\documentclass[aps,prl,twocolumn,fleqn,floatfix]{revtex4}

\usepackage{epsfig}
\usepackage{amsmath,amssymb,amsfonts}

\begin{document}

\title{Suppression of $p$-Wave Baryons in Quark Recombination}

\author{Yoshiko Kanada-En'yo}
\affiliation{Yukawa Institute for Theoretical Physics, Kyoto University,
             Kyoto 606-8502, Japan}

\author{Berndt M\"uller}
\affiliation{Yukawa Institute for Theoretical Physics, Kyoto University,
             Kyoto 606-8502, Japan}
\affiliation{Department of Physics, Duke University, Durham, NC 27708, USA}

\date{\today}

\begin{abstract}
We show that the observed suppression of the $\Lambda(1520)/\Lambda$
ratio in central Au + Au collisions at the Relativistic Heavy Ion Collider 
can be naturally understood in the constituent quark recombination 
model.
\end{abstract}

\maketitle

Recent data from nuclear collisions at the Relativistic Heavy Ion Collider 
(RHIC) show that the emission of excited hadronic states is differentially
suppressed in central Au + Au collisions \cite{Adams:2006yu,Gaudichet:2003jr}.
Similar results were also found in central collisions of Pb + Pb at the
CERN-SPS \cite{Friese:2002re}. The yields of some hadrons are suppressed in 
comparison with both, $p+p$ collisions and statistical model predictions. 
Curiously, not all excited hadrons are suppressed. For example, the ratio 
$\Sigma^*/\Lambda$ shows no suppression at RHIC, and the ratio $\phi/K$ 
appears to be even slightly enhanced. The ratio $K^*/K$ is modestly 
suppressed, and the strongest suppression is seen in the ratio 
$\Lambda^*/\Lambda$ where $\Lambda^*$ stands for the negative parity 
partner of the $\Lambda$-hyperon with a mass of 1520 MeV. 

The differences of the relative yields of these strange hadrons compared 
with yields measured in $p+p$ collisions is not very surprising, because
the production of hadrons containing strange quarks is known to be 
strongly enhanced in nuclear collisions at RHIC energies. What is 
noteworthy is that the measured yield ratios deviate from the predictions
of the thermal statistical model, which generally explains the observed 
hadron yields well \cite{Becattini:2002nv,Cleymans:2004pp,Andronic:2005yp}.
The tentative explanation for the observed deviation proposed by Adams 
{\it et al.} \cite{Adams:2006yu} is that some fraction of the short-lived 
excited hadrons decays before the medium has become so dilute that the
decay fragments can escape unscattered. It is then possible to estimate
the duration of the dense hadronic phase from the known life-time of the
hadronic state and the observed suppression \cite{Markert:2002rw}. The 
problem with this explanation is that it yields vastly different estimates 
for the duration of the dense hadronic phase. For example, the $K^*/K$ 
ratio demands a time-span $\Delta\tau\approx 2.5$ fm/c, whereas the 
$\Lambda^*/\Lambda$ ratio would require a much longer duration of 
$\Delta\tau\approx 9$ fm/c. Such a long time-span is difficult to 
reconcile with the values of the duration of the emission process 
derived from two-pion intensity interferometry measurements, which 
give values for $\Delta\tau$ in the range $2-3$ fm/c \cite{Adams:2004yc}.

The discrepancy can possibly be explained by including the effect of 
partial regeneration of excited hadronic states in the evolving dense 
medium \cite{Bleicher:2002dm}, if its life-time is sufficiently long. 
However, quantitative predictions of this effect depend on various 
unknown cross sections, and it is not clear whether a simple compelling
picture emerges that can explain all the measured ratios \cite{Adams:2006yu}. 

In this article, we propose a simple explanation for the observed 
strong suppression of the $\Lambda^*/\Lambda$ ratio, compared with the
observed $K^*/K$ ratio and the thermal model predictions. Our argument
is based on the fact that the $\Lambda(1520)$ has a different internal 
quark structure than of all other hadrons that have been detected in 
the final state of relativistic heavy ion reactions. In the framework 
of the constituent quark model the $\Lambda(1520)$ is described as an 
orbital excitation of the ground state $\Lambda$-hyperon, where one 
quark is in a $p$-wave orbital \cite{Isgur:1978xj}. In the calculations 
the $\Lambda(1520)$
is a mixture of flavor-SU(3) singlet and octet, which corresponds to a 
spatial wave function with the strange quark being predominantly in 
the $p$-wave state, specifically:
\begin{equation}
|\Lambda(1520)\rangle \approx 0.91 |1\rangle + 0.40 |8\rangle
\approx -0.37 |s\rangle + 0.93 |p\rangle \, ,
\label{eq:config}
\end{equation}
where $|1(8)\rangle$ denotes the flavor-SU(3) state and $|s(p)\rangle$
the symmetry of the internal spatial wave function of the $s$ quark. 
For simplicity, we shall neglect the minor component of the spatial 
wave function and approximate the $\Lambda^*$ state as a pure $p$-wave 
excitation of the strange quark with the $(ud)$ quark pair being in a
spin and isospin singlet state configuration. 

The internal structure of the emitted hadron is assumed to be of no 
relevance in the statistical (thermal) model of particle emission, only 
the mass and the quantum degeneracy matter. The assumption underlying the 
statistical model is that all $S$-matrix elements for hadron production
are of roughly equal magnitude. This assumption may be justified for 
the emission of ground state hadrons, which all have similar spatial 
quark wave functions. The insensitivity to the hadronic wave function 
can be demonstrated in the framework of the sudden recombination model 
in the high temperature limit, when the source is homogeneous over the
sie of the hadron \cite{Fries:2003kq}. In practice, however, the 
prevailing conditions at hadronization do not correspond to this limit.
The formation temperature of hadrons, $T_{\rm ch}\approx 160$ MeV, is 
comparable to the energy scale of hadronic excitations, and the coherence
length of quarks just before hadronization is also expected to be similar
to the size of a hadron. It is therefore plausible that details of the 
spatial wave function of the emitted hadron should have an effect on the 
production yield, leading to deviations from the predictions of the 
statistical model. 

We show below in the framework of the nonrelativistic constituent quark 
model that the internal $p$-wave structure of the $\Lambda(1520)$ 
hyperon leads to a significant suppression of its creation by quark
recombination, compared to $s$-wave baryons. For reasonable choices 
of the parameters, we find that the suppression is approximately one-half, 
similar to the experimental observation. The amount of the suppression
is quite insensitive to the parameters of our model. Our result suggests
that the observed suppression of $\Lambda(1520)$ production in central
Au + Au collision is not the consequence of final-state interactions, 
but of the production mechanism (quark recombination) itself. The insight 
that the internal $p$-wave structure of a hadron can reduce its yield 
in nuclear reactions was also reached in the framework of a quark-diquark
model of baryons \cite{Zimanyi:2004qw}. Our work differs from this earlier
study by our use of a more realistic internal wave function of the emitted
baryon, and it is based on the recombination model rather than the direct
reaction model. These qualitative differences have important quantitative
consequences, in particular, the use of a proper internal wave function of 
the baryon is essential for the magnitude of the predicted suppression.

Adopting the convention of Isgur and Karl, we denote the positions of the 
$u$- and $d$-quark by $\mathbf x_1$ and $\mathbf x_2$, respectively, and 
the position of the $s$-quark by $\mathbf x_3$. We assume that the $u$- 
and $d$-quarks have the same mass $m$, and the $s$-quark has mass $m_s$. 
The Jacobi coordinates are given by
\begin{eqnarray}
{\mathbf r}_{\rm cm} 
&=& \frac{m({\mathbf x}_1+{\mathbf x}_2)+m_s{\mathbf x}_3}{M} \, ,
\nonumber \\
{\mathbf r} &=& ({\mathbf x}_1-{\mathbf x}_2)/2 \, ,
\nonumber \\
{\mathbf r}' &=& ({\mathbf x}_1+{\mathbf x}_2)/2-{\mathbf x}_3 \, ,
\label{eq:Jacobi}
\end{eqnarray}
where $M=2m+m_s$ denotes the total mass. 
We assume that the internal wave functions of the $\Lambda$ and 
$\Lambda^*$ can be written as follows
\begin{eqnarray}
|\Lambda\rangle 
&=& 2^{-1/2} \Phi^s(b_{u/d},{\mathbf r})[\chi_u\times\chi_d]^{(0)} 
\nonumber \\    
&&  \times \Phi^s(b_s,{\mathbf r}')\chi_s^{(1/2)} |(ud-du)s\rangle \, ,
\label{eq:Lambda}
\\
|\Lambda^*\rangle 
&=& 2^{-1/2} \Phi^s(b_{u/d},{\mathbf r})[\chi_u\times\chi_d]^{(0)} 
\nonumber \\
&&  \times [\Phi^p(b_s,{\mathbf r}')\times\chi_s]^{(3/2)} |(ud-du)s\rangle \, ,
\label{eq:Lambda*}
\end{eqnarray}
where $\Phi^{s}$ and $\Phi^{p}$ are the lowest $s$- and $p$-wave 
eigenstates of the harmonic oscillator:
\begin{eqnarray}
\Phi^s(b,{\mathbf r}) &=& (\pi b^2)^{-3/4} e^{-{\mathbf r}^2/2b^2} ,
\nonumber \\
\Phi^p(b,{\mathbf r}) 
&=& (\pi b^2)^{-3/4} \frac{\sqrt{2}{\mathbf r}}{b} 
    e^{-{\mathbf r}^2/2b^2} \, .
\label{eq:hosc}
\end{eqnarray}
We take the radius parameter of the constituent $u,d$ quark wavefunctions 
as $b\approx 0.6$ fm \cite{Isgur:1979} and assume that the parameter scales as 
usual with the reduced mass: $b_{u/d}=b\sqrt{2}$, $b_s=b\sqrt{M/2m_s}$.

\begin{figure}
\centerline{\includegraphics[width=\linewidth]{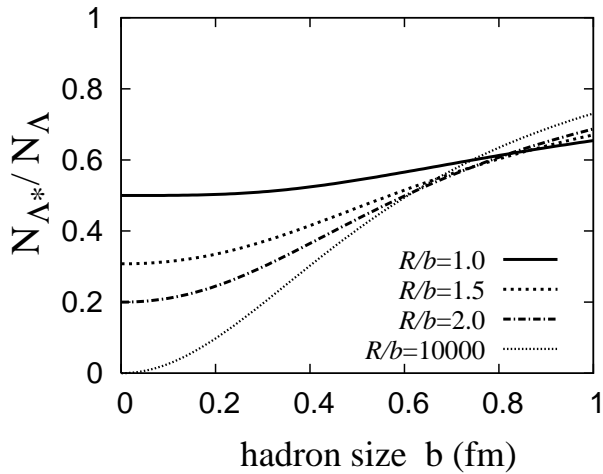}}
\caption{Suppression factor for the formation of a $p$-wave baryon by
recombination of three quarks from a thermal medium. The suppression 
factor is shown as a function of the hadron size parameter $B$ for three
different choices of the coherence length $R$ of the initial-state quarks
($R/b=1,1.5,2,10^4$). The temperature and the $ud$-quark mass are $T=160$ 
MeV and $m=330$ MeV, and the hadronization surface is assumed to have a 
thickness $a=0.5$ fm.}
\label{f:ratio_vs_b}
\end{figure}

We further assume that the quarks just before hadronization can be 
described by Gaussian wave packets of the form
\begin{equation}
\Psi(R,{\mathbf D},{\mathbf k};{\mathbf x}) = (\pi R^2)^{-3/4} 
  \exp\left(-\frac{1}{2R^2}({\mathbf r}-{\mathbf D})^2
            +i{\mathbf k}\cdot{\mathbf r}\right) \, ,
\label{eq:Gauss}
\end{equation}
where $\mathbf D$ denotes the position and $\mathbf k$ the momentum of 
the center of the wave packet. The width parameter $R$ describes the
coherence properties of the initial quarks, which is determined by their
mean free path in the hadronizing medium. Such Gaussian wave packets 
have been used, e.~g., in the framework of quark molecular dynamics 
\cite{Maruyama}. Because the magnitude of $R$ is not well known from
phenomenological considerations, we shall explore a range of values 
for $R$ in the vicinity of the confinement radius. We allow $R^2$ to 
scale inversely with the quark mass in order to ensure that quarks become
localized in the limit of infinite mass. We further assume that the 
initial quark momenta $\mathbf k$ are given by a thermal distribution 
with temperature $T=T_{\rm ch}$. Finally, we assume that the hadrons 
are formed by recombination in a surface region, which we model as a 
plane with a Gaussian thickness profile of half-width $a$. The ensemble
distribution for the initial state wave packet thus has the form
\begin{equation}
F({\mathbf D},{\mathbf k}) 
= \prod_{i=1}^3 \exp\left(-\frac{D_{i,x}^2}{2a^2} 
                          -\frac{{\mathbf k}_i^2}{2m_iT}\right) \, .
\label{eq:FDk}
\end{equation}
With these model assumptions, the (sudden) transition matrix elements
from an uncorrelated three-quark state to the hadronic states 
(\ref{eq:Lambda}) and (\ref{eq:Lambda*}) can be easily evaluated. 
The thermal average and the integration over the hadronization zone 
can also be performed analytically. We denote the averaged transition
probabilities as $W^{(s)}$ and $W^{(p)}$, respectively:
\begin{equation}
W^{(s)} 
= \int \prod_i d{\mathbf D}_i d{\mathbf k}_i F({\mathbf D},{\mathbf k}) 
  \left|\langle\Lambda|\Psi_i\rangle\right|^2 \, ,
\label{eq:prob}
\end{equation}
and the analogous equation giving $W^{(p)}$ for $\Lambda^*$ formation. 
The final result for the ratio of the transition probabilities into the
$\Lambda$ and $\Lambda^*$ states, correcting for their 
spin$(J_\Lambda,J_{\Lambda^*})$ degeneracies, 
is:
\begin{widetext}
\begin{equation}
\frac{(2J_{\Lambda}+1)W^{(p)}}{(2J_{\Lambda^*}+1)W^{(s)}} 
= \frac{2b^2}{3(R^2+b^2)} 
  \left[ 1 + \left(2+\frac{R^2+b^2}{a^2}\right)^{-1} 
  + 3R^4\left(2b^2R^2+\frac{R^2+b^2}{mT}\right)^{-1} \right] \, .
\label{eq:ratio}
\end{equation}
\end{widetext}
We first note that the ratio tends to one in the limit $R,T\to\infty$.
This is not unexpected as it confirms the result obtained by Fries {\em 
et al.} \cite{Fries:2003kq}. For finite values of $R$ and $T$ the ratio
is always less than one, i.~e.\ the production of the $p$-wave hadron 
(the $\Lambda^*$) is suppressed. This is explored in Fig.~\ref{f:ratio_vs_b}, 
which shows the suppression factor (\ref{eq:ratio}) as a function of
the hadronic size parameter $b$ for several different choices of the 
quark coherence length $R$. Remarkably, in the range of realistic values
for $b\approx 0.6$ fm) the suppression is insensitive to the precise value
of $R$. This is gratifying, because $R$ is by far the most uncertain 
parameter in our model. In the large-$a$ limit for the surface thickness,
which corresponds to a volume-dominated quark recombination, the suppression
factor is found to be slightly (about 10\%) enhanced. 

Figure \ref{f:ratio_vs_b} shows that the formation of $p$-wave baryon
is suppressed by a factor $0.5-0.6$ in the realistic range of parameters
in our model. This is quite close to the suppression, compared with the
statistical model prediction, of the $\Lambda(1520)/\Lambda$ ratio in 
central Au + Au collisions \cite{Adams:2006yu}. We do not need to adjust 
the hadron size parameter $b$ to fit the observed suppression value, 
because we are using a realistic hadron wave function. We emphasize that 
this supprssion 
is a formation effect, not an effect of final-state interactions. Within
our model it is predicted that the suppression should be independent of 
the momentum of the baryon. If high quality spectra of $\Lambda$ and 
$\Lambda^*$ hyperons can be obtained in future high statistics runs, 
this property could perhaps be used to distinguish between the suppression 
mechanism proposed here and final-state suppression mechanisms. On general 
grounds one expects that a suppression effect due to rescattering will 
diminish with increasing hadron momentum, because the particle spends less 
time in the medium. Of course, this argument applies only to the momentum
range in which recombination is thought to be the dominant hadronization 
process ($p_T \leq 4$ GeV/c). Also, relativistic effects may modify our
prediction.

In conclusion, we have shown that under the conditions prevailing at 
hadronization of a quark-gluon plasma the recombination model predicts 
that the yield of hadrons with an internal $p$-wave excitation is 
suppressed compared with the yield of ground state hadrons. For the 
example of $\Lambda(1520)$ versus $\Lambda$ formation, the suppression
factor is calculated to be in the range $0.5-0.6$. This prediction is 
found to be rather insensitive to the precise choice of parameters of 
the model. We suggest the mechanism pointed out here may be the origin 
of the relative suppression, compared to statistical model expectations, 
of $\Lambda(1520)$ production in central Au + Au collisions at RHIC.

The present work is based on the idea of the recombination model. It 
extends previous work by taking into account the internal structure of
the emitted baryon and obtains the new insight that an internal $p$-wave
may result in a reduced hadron yield. We note that the same mechanism 
affects the production of all $p$-wave hadrons. It would be interesting 
to confirm this experimentally, but unfortunately is seems unlikely that 
the $p$-wave excitation of any other baryon or meson can be identified 
in the final state of a heavy ion collision.

{\it Acknowledgments:} B.~M.\ would like to thank the members of the 
YITP, especially T.~Kunihiro, for their hospitality during his stay in 
Kyoto as YITP visiting professor. Y.~K-E. thanks T.~Kunihiro and 
H.~Tsunetsugu for valuable discussions. This work was supported in 
part by the Japan Society for the Promotion of Science and a 
Grant-in-Aid for Scientific Research from the Japan Ministry of 
Education, Science and Culture.

\end{document}